\def\year{2021}\relax
\documentclass[letterpaper]{article} 
\usepackage{aaai21}  
\usepackage{times}  
\usepackage{helvet} 
\usepackage{courier}  
\usepackage[hyphens]{url}  
\usepackage{graphicx} 
\usepackage{xcolor}
\usepackage{natbib}  
\usepackage{caption} 
\usepackage{amsmath}
\usepackage{float}
\usepackage{multirow}

\title{Reduced-order Model for Fluid Flows via Neural Ordinary Differential Equations}
\author {
    Carlos J.G. Rojas \textsuperscript{\rm 1},
    Andreas Dengel \textsuperscript{\rm 1},  
    Mateus Dias Ribeiro \textsuperscript{\rm 1} \\
}
\affiliations {
    \textsuperscript{\rm 1} German Research Center for Artificial Intelligence - DFKI\\
    {carlos.gonzalez\_rojas@dfki.de, andreas.dengel@dfki.de, mateus.dias\_ribeiro@dfki.de}
}

\begin{document}

\maketitle

\begin{abstract}

Reduced order models play an important role in the design, optimization and control of dynamical systems. In recent years, there has been an increasing interest in the application of data-driven techniques for model reduction that can decrease the computational burden of numerical solutions, while preserving the most important features of complex physical problems. In this paper, we use the proper orthogonal decomposition to reduce the dimensionality of the model and introduce a novel generative neural ODE (NODE) architecture to forecast the behavior of the temporal coefficients. With this methodology, we replace the  classical Galerkin projection with an architecture characterized by the use of a continuous latent space. We exemplify the methodology on the dynamics of the Von Karman vortex street of the flow past a cylinder generated by a Large-eddy Simulation (LES)-based code.  We compare the NODE methodology with an LSTM baseline to assess the extrapolation capabilities of the generative model and present some qualitative evaluations of the flow reconstructions. 

\end{abstract}

\section{Introduction}

Modeling and simulation of dynamical systems are essential tools in the study of complex  phenomena with applications in chemistry, biology, physics and engineering, among other relevant fields. These tools are particularly useful in the control and design of parametrized systems in which the dependence on properties, initial conditions and other configurations requires multiple evaluations of the system response. However, there are some limitations when performing numerical simulations of systems where nonlinearities, and  a wide range of spatial and time scales leads to unmanageable demands on computational resources. The latter is the case of engineering fluid flow problems where the range of scales involved increase with the value of the Reynolds number and the cost of simulating a full-order model (FOM) using techniques such as DNS or LES is very high.  One of the possible solutions to reduce the expensive computational cost is to introduce an alternative, cheaper and faster representation that retains the characteristics provided by the FOM without sacrificing the accuracy of the general physical behaviour. The idea is to construct a methodology able to generalize the physical behaviour for unseen parameters and that can extrapolate forward in time using the minimal amount of full order simulations \cite{benner_survey_2015}.

The projection-based reduced order modeling is one of the most popular approaches to construct surrogate models of dynamical systems. This framework reduces the degrees of freedom of the numerical simulations using a transformation into a suitable low-dimensional subspace. Then, the state variable in the governing equations is rewritten in terms of the reduced subspace and finally the PDE equations are converted into a system of ODEs that can be solved using classical numerical techniques \cite{benner_survey_2015}. In the field of fluid mechanics, the Proper Orthogonal Decomposition (POD) method is widely applied in the dimensionality reduction of the FOM and the Galerkin method is used for the projection onto the governing equations. These methodologies are preferred because an orthogonal normal basis simplifies the complexity of the projected mathematical operators and the truncated basis of the POD is optimal in the least squares sense, retaining the dominant behaviour through the most energetic modes. The projection on the governing equations maintain the physical structure of the model, but the truncation of the modes can affect the accuracy of the results in nonlinear systems and it may also be restricted to stationary and time periodic problems. Furthermore, the projection is intrusive, requiring different settings for each problem, and it is limited to explicit and closed definitions of the mathematical models \cite{san_artificial_2019}. Some of these problems have been addressed with the search of closure models that compensates the information losses produced by the truncated modes \cite{mou_data-driven_2020,mohebujjaman_physically_2019,san_neural_2018, san_machine_2018} and with the construction of a data driven reduced "basis" that  also provides optimality after the time evolution \cite{ murata_nonlinear_2020, liu_novel_2019,wang_deep_2016}. 

\begin{figure*}[hbt!]
\centering
\includegraphics[width=1.0\textwidth]{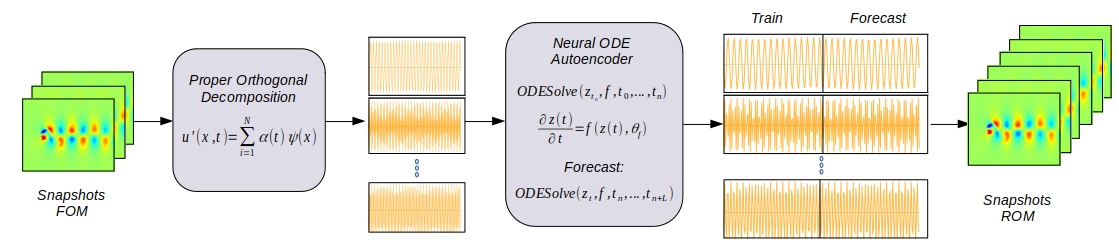} 
\caption{POD-NeuralODE ROM methodology.}
\label{fig1}
\end{figure*}

We present an alternative methodology to evolve the dynamics of the system in the reduced space using a data-driven approach. We use the POD to compute the modes and the temporal coefficients of a fluid flow simulation and then we apply a non-supervised autoencoder approach to learn the dynamics of a latent space. The addition of a neural ODE \cite{chen_neural_2019, rubanova2019latent} block in the middle of the autoencoder model provides a continuous learning block that is encoded using a feed forward neural network and that can be solved numerically to determine the future states of the input variables. Several works have proposed machine learning models to replace the Galerkin projection step or to improve their capabilities, and different architectures such as feedforward or recurrent networks has been applied with demonstrated good performance in academic and practical fluid flow problems \cite{pawar_deep_2019,imtiaz_pod-based_2020,eivazi_deep_2020,lui_construction_2019,portwood2019turbulence,maulik_probabilistic_2020,maulik_time-series_2020}. The main advantage of the neural ODE generative model is that the learning is posed as a non-supervised task using a continuous representation of the physical behavior. In our view, the neural ODE block can be interpreted as an implicit differential operator that is not restricted to a specific differential equation. This setting provides more flexibility than the projection over the governing equations because it addresses the learning problem with an operator that is informed and corrected by the training data.

\section{Methodology}

In this work we use a Large-eddy Simulation (LES) model to approximate the behavior of the fluid flow dynamical system. As it is the case in many fluid flow problems, the discrete solution has a spatial dimension larger than the size of the temporal domain.  For this reason, we apply the snapshot POD to construct the reduced order model in order to have a tractable computation. The POD finds a new basis representation that maximizes the variance in the data, and has the minimum error of the reconstructions in a least squares sense. In addition, the dimensionality reduction is easily performed because the components of the new basis are ordered by their contribution to the recovery of the data. 

The main block in the Neural ODE-ROM methodology is concerned with the forecast of the temporal coefficients provided by the snapshot POD. Here we apply a generative neural ODE model that takes the temporal coefficients, learns their dynamical evolution and  provides an adequate model to extrapolate at the desired time steps.  Finally, we can forecast the evolution of the temporal coefficients and reconstruct the behavior of the flow with the spatio-temporal expansion used in the POD.

The general methodology is represented in the Fig.~\ref{fig1} and more details  about each of the building blocks is presented in the following sections.

\section{LES Model For The Flow Past a Cylinder}


The dynamics of the Von Karman vortex street of the flow past a cylinder were solved by the LES filtered governing equations for the balance of mass (\ref{eq:ns1}), and momentum (2), which can be written as follows: 

\begin{equation}
\label{eq:ns1}
\frac{\partial \bar{\rho}}{\partial t} + \frac{\partial (\bar{\rho}\tilde{u_i})}{\partial x_i} = 0
\end{equation}

\begin{equation}
\begin{split}
\label{eq:ns2}
\frac{\partial (\bar{\rho} \tilde{u_i})}{\partial t} + \frac{\partial (\bar{\rho} \tilde{u_i} \tilde{u_j})}{\partial x_j} = \frac{\partial}{\partial x_j} \bigg[ \bar{\rho} \bar{\nu} \bigg( \frac{\partial \tilde{u_j}}{\partial x_i} + \frac{\partial \tilde{u_i}}{\partial x_j} \bigg) \\- \frac{2}{3}\bar{\rho}\bar{\nu}\frac{\partial \tilde{u_k}}{\partial x_k}\delta_{ij} - \bar{\rho}{\tau_{ij}}^{sgs}  \bigg] - \frac{\partial \bar{p}}{\partial x_i} + \bar{\rho}g_i
\end{split}
\end{equation}

In the previous equations, $u$ represents the velocity, $\mathrm{\rho}$ is the fluid density, and $\mathrm{\nu}$ is the dynamic viscosity. These equations are solved numerically using the PIMPLE algorithm \cite{openfoam}, which is a combination of PISO (Pressure Implicit with Splitting of Operator) by \citet{issa} and SIMPLE (Semi-Implicit Method for Pressure-Linked Equations) by \citet{patankar}. This approach obtains the transient solution of the coupled velocity and pressure fields by applying the SIMPLE (steady-state) procedure for every single time step. Once converged, a numerical time integration scheme (e.g. backward) and the PISO procedure are used to advance in time until the simulation is complete. Furthermore, the unresolved subgrid stresses, ${\tau_{ij}}^{sgs}$, are modeled in terms of the subgrid-scale eddy viscosity $\nu_T$ using the dynamic $k$-equation approach by \citet{menon}.









\begin{figure}[H]
\centering
\includegraphics[width=0.54\columnwidth]{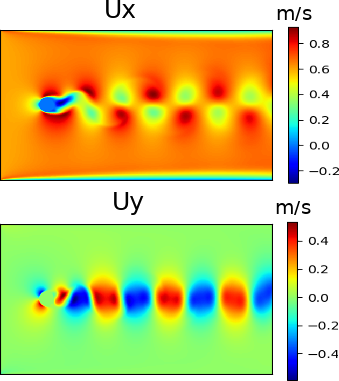} 
\caption{Snapshot of the flow field at t = 100.}
\label{fig2}
\end{figure}

The setup of the problem is described as follows. The computational domain comprehends a 2D channel with 760~mm in the stream-wise direction and 260~mm in the direction perpendicular to the flow. The cylinder is located between the upper and bottom walls of the channel at 115~mm away from the inlet (left wall). A constant radial velocity of 0.6 m/s with random radial/vertical fluctuations in combination with a zero-gradient outflow condition and non-slip walls on the top/bottom/cylinder walls are imposed as boundary conditions. Furthermore, a laminar dynamic viscosity of $1\times10^{-4} m^2/s$ and a cylinder diameter of 40~mm further characterizes the flow with a Reynolds number of 240 ($Re = 0.6\times0.04/1\times10^{-4} = 240$). The Central differencing scheme (CDS) was used for the discretization of both convective and diffusive terms of the momentum equation, as well as an implicit backward scheme for time integration. A snapshot of the velocity components in both radial and axial directions at time = 100 is shown in the Fig.~\ref{fig2}.

\section{Proper Orthogonal Decomposition}

The proper orthogonal decomposition (POD) is known under a variety of names such as  Karhunen-Loeve expansion, Hotelling transform  and principal component analysis \cite{liang_proper_2002}. This tool was developed in the field of probability theory to discover interdependencies within vector data and introduced in the fluid mechanics community by \citet{berkooz_proper_1993}.  Once the interdependencies in the data are discovered, it is possible to reduce its dimensionality.

The formulation of the dimensionality reduction starts with some samples of  observations provided by experimental results or obtained through the numerical solution of a full order model that characterizes the physical problem. These samples are rearranged in an ensemble matrix of snapshots  $Y$ where each row has the state of the dynamical system at a given time step.  Then, the correlation matrix of the elements in  $Y$ is computed and their eigenvectors are used as an orthogonal optimal new basis for the reduced space. 

In the following list we summarize the main steps used for the construction of the snapshot POD:

\begin{itemize}
\item Take snapshots : simulate the dynamical system and sample its state as it evolves. 
\item Compute the fluctuating components of the velocity using the Reynolds decomposition of the flow:

 \begin{equation}
            u = \overline{u} + u^{\prime},
        \end{equation}
where $\overline{u}$ is the temporal mean of the solutions given by the FOM model.

\item Assemble the matrix $Y$ with the snapshots in the following form: 
\begin{center}
$ Y = \begin{bmatrix}
\tiny
u_x^{\prime}(x_1, y_1, t_1) & ... & u_y^{\prime}(x_{N_x}, y_{N_y}, t_1) \\
u_x^{\prime}(x_1, y_1, t_2) & ... & u_y^{\prime}(x_{N_x}, y_{N_y}, t_2) \\
.&.&.\\
.&.&.\\
.&.&.\\
u_x^{\prime}(x_1, y_1, t_{N_t}) & ... & u_y^{\prime}(x_{N_x}, y_{N_y}, t_{N_t}) \\
\end{bmatrix}$
\end{center}

where each row contains a flattened array with the fluctuating components of the velocity in the $x$ and $y$ directions for a given time step.  If the discretization used for the FOM simulation has dimensions $N_x, N_y$ and $N_t$,  then the flattened representation is a vector with length $2\cdot N_x \cdot N_y$  and the matrix $Y$ has dimensions $  N_t \times (2\cdot N_x \cdot N_y)  $.

\item Build the correlation matrix $K$ and compute its eigenvectors $a_j$: 

 \begin{equation}
            K = Y Y^{\top},
\end{equation}

 \begin{equation}
            K a_j = \lambda_j a_j.
\end{equation}

Alternatively, one can directly compute the eigenvalues and eigenvectors using the singular value decomposition (SVD) of the snapshot matrix. 

\item Choose the reduced dimension of the model:  As described in the literature, larger eigenvalues are directly related with the dominant characteristics of the dynamical system while small eigenvalues are associated with perturbations of the dynamic behavior. The criterion to select the components  for the new basis is to maximize the relative information content $I(N)$ using the minimal amount of components $N$ necessary to achieve a desired percentage of recovery \cite{schilders_model_2008}.

 \begin{equation}
            I(N) = \frac{\sum_{i=1}^{N} \lambda_i}{\sum_{i=1}^{N_t} \lambda_i}
\end{equation}

\item Finally, we compute the spatial modes using the temporal coefficients in the reduced dimensional space and the Ansatz decomposition of the POD:

\begin{equation}
            u^{\prime}=  \sum_{i=1}^{N} \alpha_i(t) \psi_i(x),
\end{equation}

\begin{equation}
            \psi_i(x)=  \frac{1}{\lambda_i} \sum_{j=1}^{N} \alpha_i(t_j)  u^{\prime}(t_j).
\end{equation}

\end{itemize}

\section{Neural Ordinary Differential Equations}

The neural ordinary differential methodology \cite{chen_neural_2019} can be interpreted as a continuous counterpart of traditional models such as recurrent or residual neural networks. In order to formulate this model, the authors drew a parallel between the classical composition of a sequence in terms of previous states and the discretization methods used to solve differential equations: 

  \begin{equation}
    h_{t+1} =  h_t + f(h_t , \theta).
 \label{Eq:1}
 \end{equation} 

In the limit case of sufficient small steps (equivalent to an increase of the layers) is possible to write a continuous parametrization of the hidden state derivative:

\begin{equation}
 \frac{dh(t)}{dt} = f(h_t , \theta), 
 \end{equation} 

\begin{figure*}[ht]    
\centering
\includegraphics[width=0.8\textwidth]{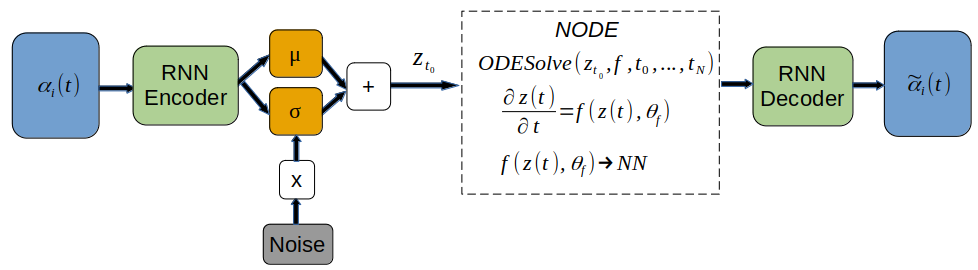} 
\caption{Generative VAE with Neural ODE.}
\label{fig3}
\end{figure*}

  \begin{equation}
   h_t = ODESolver(h_0 , f(h_t , \theta)).
 \end{equation} 

The function $f$ defining the parametrization of the derivative can be approximated using a neural network and the values of hidden states $h_t$ at different time steps are computed using numerical ODE solvers \cite{chen_neural_2019}.

We apply the neural ODE time-series generative approach presented in the Fig.~\ref{fig3} to model the evolution of the temporal modes provided by the proper orthogonal decomposition. This approach can be interpreted as a variational autoencoder architecture with an additional neural ODE block after the sampling of the codings. This block maps the vector of the initial latent state $z_{t_0}$ to a sequence of latent trajectories using the ODE numerical solver while a neural network $f(h_t , \theta)$ learns the latent dynamics necessary to have  a good reconstruction of the input data. 

After the training process, the latent trajectories are easily extrapolated  with the redefinition of the temporal bounds in the ODE solver. Some of the advantages of this strategy are that it does not need an explicit formulation of the physical laws to forecast the temporal modes, and in consequence, the method does not resort onto projection methodologies. Furthermore, the parametrization using a neural network gives an accurate nonlinear approximation of the derivative without a predefined mathematical structure.

\section{Results}

In this section, we evaluate the performance of the generative neural ODE model in the forecasting of the temporal coefficients. For this assessment, we apply the proper orthogonal decomposition over 300 snapshots of simulated data obtained with the LES code and take the first 8 POD modes achieving a 99 $\%$ of recovery according to the relative information content. For the deployment of the neural ODE model (NODE) we take the first 75 time steps for the training set, the following 25 time steps for the validation of the model and the last 200 time steps for the test set. Furthermore, we employ as a baseline model an LSTM sequence to vector architecture as proposed in  \citeauthor{maulik_time-series_2020} with a window size of 10 time steps and a batch size of 15 sequences. 

We tuned the hyperparameters necessary for both models adopting a random search and chose the best configuration given the performance on the validation set. The evolution of the loss for the best model is shown in Fig.~\ref{fig4} and the set of hyperparameters employed are presented in Table \ref{tab1}. 

\begin{table}[t]
\begin{tabular}{ccccll}
\cline{1-4}
\multicolumn{1}{|c|}{Model}                       & \multicolumn{1}{c|}{Hyperparameter}   & \multicolumn{1}{c|}{Range}       & \multicolumn{1}{c|}{Best}   &                      &  \\ \cline{1-4}
\multicolumn{1}{|c|}{\multirow{7}{*}{Neural ODE}} & \multicolumn{1}{c|}{latent dimension} & \multicolumn{1}{c|}{3-6}         & \multicolumn{1}{c|}{6}      &                      &  \\
\multicolumn{1}{|c|}{}                            & \multicolumn{1}{c|}{layers encoder}   & \multicolumn{1}{c|}{1-5}         & \multicolumn{1}{c|}{4}      &                      &  \\
\multicolumn{1}{|c|}{}                            & \multicolumn{1}{c|}{units encoder}    & \multicolumn{1}{c|}{10-50}       & \multicolumn{1}{c|}{10}     &                      &  \\
\multicolumn{1}{|c|}{}                            & \multicolumn{1}{c|}{units node}       & \multicolumn{1}{c|}{10-50}       & \multicolumn{1}{c|}{12}     &                      &  \\
\multicolumn{1}{|c|}{}                            & \multicolumn{1}{c|}{layers decoder}   & \multicolumn{1}{c|}{1-5}         & \multicolumn{1}{c|}{4}      &                      &  \\
\multicolumn{1}{|c|}{}                            & \multicolumn{1}{c|}{units decoder}    & \multicolumn{1}{c|}{10-50}       & \multicolumn{1}{c|}{41}     &                      &  \\
\multicolumn{1}{|c|}{}                            & \multicolumn{1}{c|}{learning rate}    & \multicolumn{1}{c|}{0.001 - 0.1} & \multicolumn{1}{c|}{0.0015} &                      &  \\ \cline{1-4}
\multicolumn{1}{|c|}{\multirow{3}{*}{LSTM}}       & \multicolumn{1}{c|}{layers}           & \multicolumn{1}{c|}{10-60}       & \multicolumn{1}{c|}{49}     &                      &  \\
\multicolumn{1}{|c|}{}                            & \multicolumn{1}{c|}{units}            & \multicolumn{1}{c|}{1-5}         & \multicolumn{1}{c|}{1}      & \multicolumn{1}{c}{} &  \\
\multicolumn{1}{|c|}{}                            & \multicolumn{1}{c|}{learning rate}    & \multicolumn{1}{c|}{0.001 - 0.1} & \multicolumn{1}{c|}{0.0081} & \multicolumn{1}{c}{} &  \\ \cline{1-4}
\end{tabular}
\caption{Hyperparameters used in the models.}
\label{tab1}
\end{table}

\begin{figure}[H]
\centering
\includegraphics[width=0.9\columnwidth]{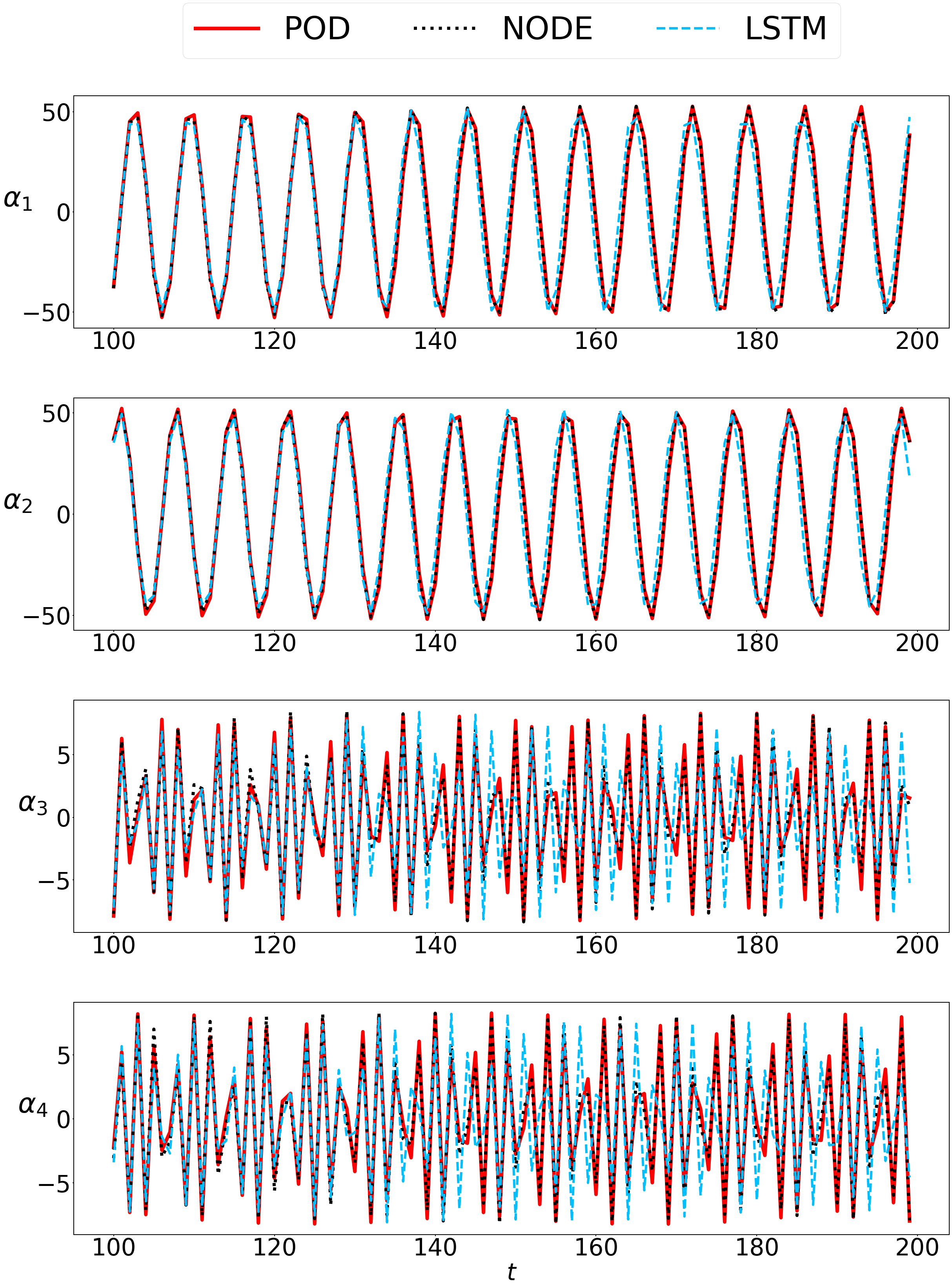} 
\caption{Reconstruction of POD temporal coefficients using NODE vs LSTM , $t \in [100,200]$.}
\label{fig5}
\end{figure}

\begin{figure}[H]
\centering
\includegraphics[width=0.9\columnwidth]{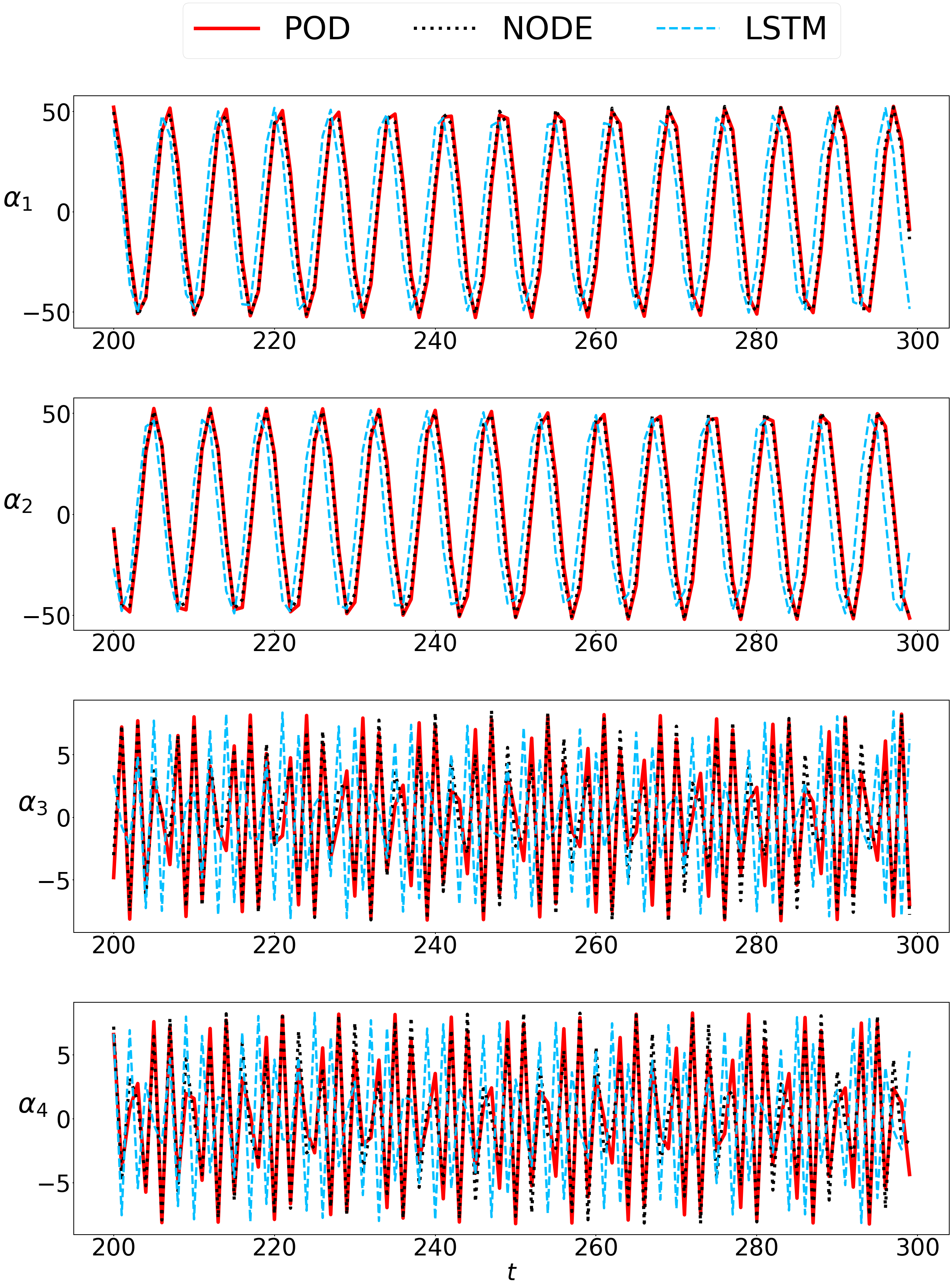} 
\caption{Reconstruction of POD temporal coefficients using NODE vs LSTM, $t \in [200,300]$.}
\label{fig52}
\end{figure}

\begin{figure}[h]
\centering
\includegraphics[width=0.9\columnwidth]{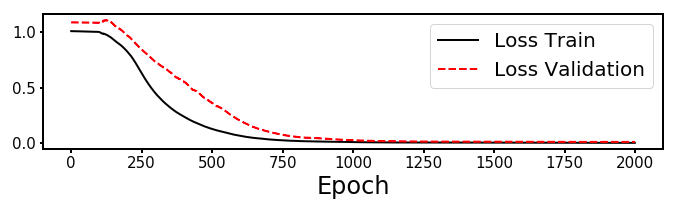} 
\caption{Loss Generative NODE model.}
\label{fig4}
\end{figure}

The time-series prediction for the first four temporal coefficients in the test set is shown in Fig.~\ref{fig5}. This plot presents the ground truth values of the POD time coefficients, the baseline produced  using an LSTM architecture and the predictions by the proposed generative NODE model for the first 100 time steps in the test window. We notice that the baseline and the NODE model learned adequately the evolution of the two most dominant coefficients, but the performance of the NODE model is significantly better for the third and four time coefficients. Additionally, the quality of the prediction using the LSTM model for the last 100 time steps in the test set deteriorates with the evolution of the time steps even for $\alpha_1$ and $\alpha_2$ as seen in Fig.~\ref{fig52} .  One of the possible reasons for this is that the autoregressive nature of the predictions in the LSTM model is prone to the accumulation of errors as \citeauthor{maulik_time-series_2020} pointed out in their study \cite{maulik_time-series_2020}. 

\begin{figure}[H]
\centering
\includegraphics[width=0.9\columnwidth]{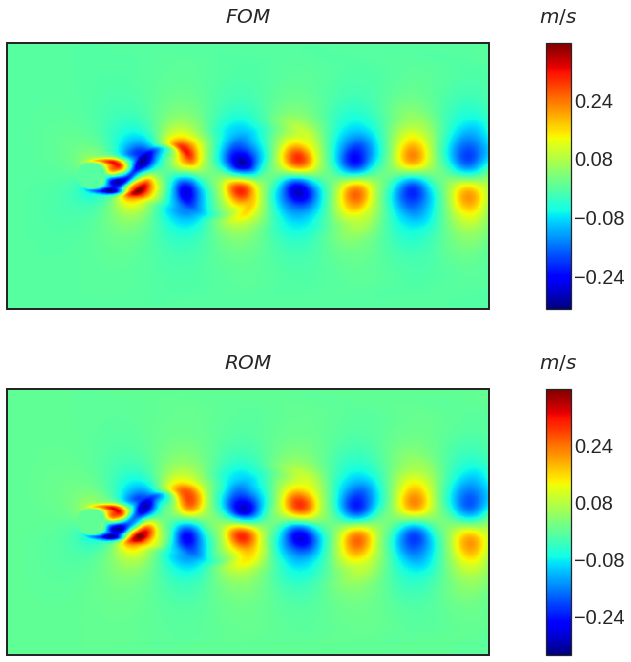} 
\caption{Contours of fluctuating component ${u_x^\prime}$, $t = 300$.}
\label{fig6}
\end{figure}

\begin{figure}[H]
\centering
\includegraphics[width=0.9\columnwidth]{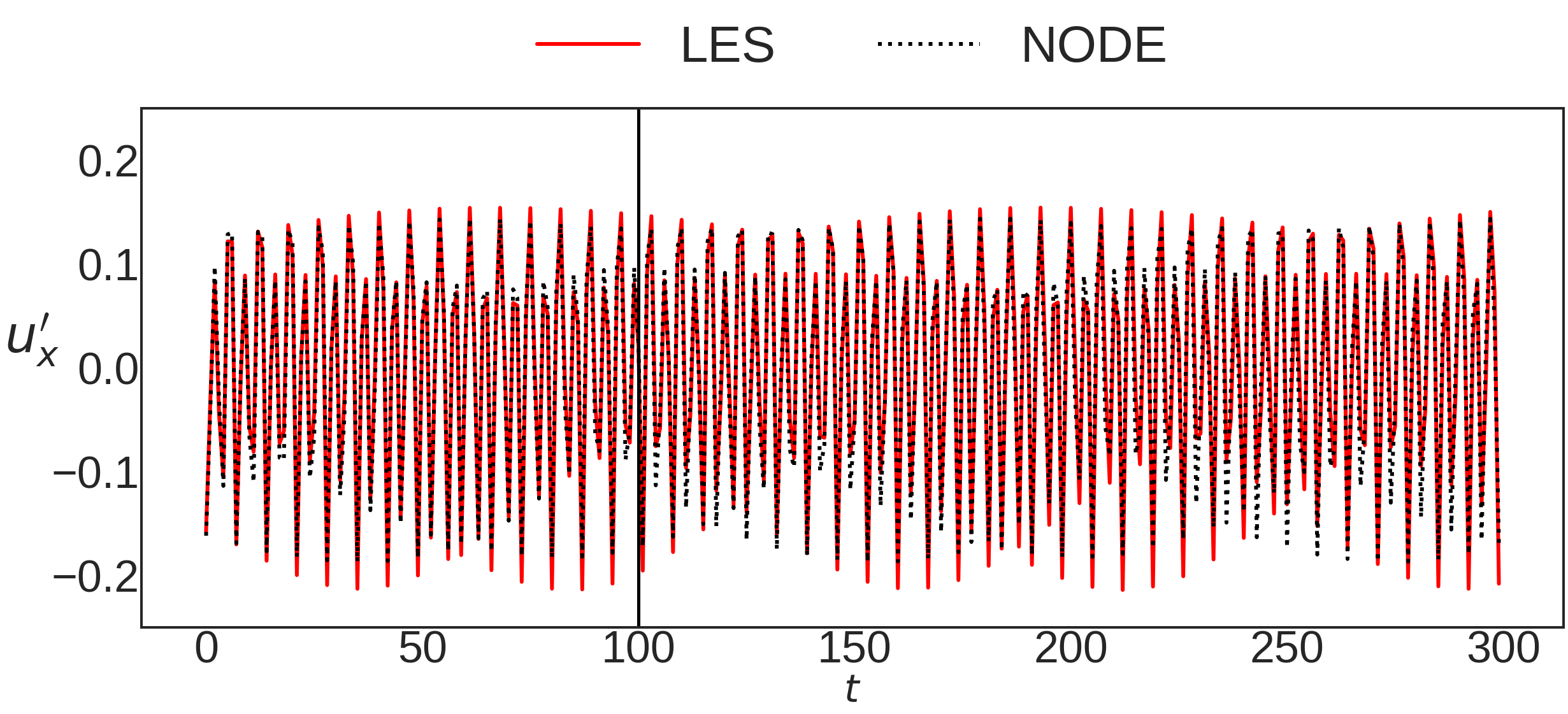} 
\caption{Probe positioned behind the cylinder.}
\label{fig7}
\end{figure}

 After the training and validation process, we reconstruct the velocity fluctuating component $u_x^{\prime}$ using the Ansatz of the proper orthogonal decomposition with the temporal coefficients forecasted for the test set. Observing the Fig.~\ref{fig6} is possible to notice that the contour generated with the reduced order model provides an adequate recovery of the flow features with only slight differences in some vortexes. In addition, we also present the fluctuation time history for a probe located downstream from the cylinder in Fig.~\ref{fig7}. This figure shows with more details how the physical response of the reduced order model gives a satisfactory approximation of the flow behavior.   

The data and code that support this study is provided at   \url{https://github.com/CarlosJose126/NeuralODE-ROM}.

\section{Conclusions}

We presented a methodology to produce reduced order models using a neural ODE generative architecture for the evolution of the temporal coefficients. The neural ODE model was able to learn appropriately the hidden dynamics of the temporal coefficients without having the propagation of errors common in the autoregressive architectures. Another advantage of this methodology is that the learning is posed as an unsupervised task without the requirement to divide the whole sequence in smaller training windows with labels. We also remark that the continuous nature of the neural ODE block is crucial for the good extrapolation capabilities of the methodology.  Finally, we expect to test the capabilities of this methodology with other physical problems and also to extend the method for parametric dynamical systems.

\bibliography{references.bib}

\end{document}